\documentclass[journal]{IEEEtran}
\usepackage{lineno}
\usepackage{amssymb}
\usepackage{graphicx}
\usepackage{booktabs} 
\usepackage{algorithm}
\usepackage{algorithmicx} 
\usepackage{algcompatible}
\usepackage{amsmath,bm}
\usepackage{multirow}
\usepackage{array}
\usepackage{subfigure}
\usepackage[justification=centering]{caption}
\usepackage{epstopdf}
\usepackage{float}
\usepackage{graphicx}
\usepackage{adjustbox}
\graphicspath{{./Figures/}}
\usepackage{mathtools}
\usepackage[noadjust]{cite}
\usepackage{filecontents}
\usepackage[noend]{algpseudocode}
\usepackage{algpseudocode}

\begin{document}
\title{Proactive Video Chunks Caching and Processing for Latency and Cost Minimization in Edge Networks}
\author{\IEEEauthorblockN{Emna Baccour\IEEEauthorrefmark{1},
		Aiman Erbad\IEEEauthorrefmark{1},
		Amr Mohamed\IEEEauthorrefmark{1},
		Kashif Bilal\IEEEauthorrefmark{2}, and
		Mohsen Guizani\IEEEauthorrefmark{1}\\
	\IEEEauthorblockA{\IEEEauthorrefmark{1}CSE department, College of Engineering, Qatar University.}\\
	\IEEEauthorblockA{\IEEEauthorrefmark{2}COMSATS Institute of Information Technology, Pakistan.} }}
\maketitle
\begin{abstract}
Recently, the growing demand for rich multimedia content such as Video on Demand (VoD) has made the data transmission from content delivery networks (CDN) to end-users quite challenging. Edge networks have been proposed as an extension to CDN networks to alleviate this excessive data transfer through caching and to delegate the computation tasks to edge servers. To maximize the caching efficiency in the edge networks, different Mobile Edge Computing (MEC) servers assist each others to effectively select which content to store and the appropriate computation tasks to process. In this paper, we adopt a collaborative caching and transcoding model for VoD in MEC networks. However, unlike other models in the literature, different chunks of the same video are not fetched and cached in the same MEC server. Instead, neighboring servers will collaborate to store and transcode different video chunks and consequently optimize the limited resources usage. Since we are dealing with chunks caching and processing, we propose to maximize the edge efficiency by studying the viewers watching pattern and designing a probabilistic model where chunks popularities are evaluated. Based on this model, popularity-aware policies, namely Proactive caching policy (PcP) and Cache replacement Policy (CrP), are introduced to cache only highest probably requested chunks. In addition to PcP and CrP, an online algorithm (PCCP) is proposed to schedule the collaborative caching and processing. The evaluation results prove that our model and policies give better performance than approaches using conventional replacement policies. This improvement reaches up to 50\% in some cases.
\end{abstract}
\section {Introduction}
In the last decade, the video content traffic witnessed an explosive growing, especially with the upgrade of the next generation mobile networks and the advancement of smart devices. For example, authors in \cite{statistics2016} stated that, in 2016, audio and video streaming contents presented the largest traffic category. This traffic is accounted for 60\% of the overall data traffic and it is predicted to increase to 78 \% by 2021 \cite{mobileData}. Notably, the huge multimedia traffic load is caused by the redundant delivery of the same popular videos. 

In order to solve the problem of repeated video transmission and support the continuously growing content delivery, MEC networks have been introduced to complement the cloud and CDN networks. In fact, in MEC networks, the base stations (BSs) are equipped with servers having small caching and computing capacities. These resources allow the network to fetch a content from the CDN only one time, then, cache and serve it in the proximity of viewers without duplicating the transmission. Since the MEC network presents an opportunity to enhance the Quality of experience (QoE) of mobile users and alleviate the data load over the transit links, it arouses the research interest. The authors in \cite{cache1} proposed a \textit{CachePro} approach where they used the storage and computing capabilities of one MEC server to store videos at the edge base stations. However, since one MEC server has a limited storage and computation capacity, collaborative and intelligent caching are introduced. In \textit{CoCache} \cite {cache5}, authors implemented video sharing to minimize the network cost while authors of \textit{JCCP} approach \cite {joint2} implemented the Adaptive Bitrate streaming technology (ABR) to jointly transcode and share new bitrate versions of videos. In the existing literature, authors propose to bring the requested long video from the cloud and to store it or transcode it in one of the MEC servers. 

However, in real life scenarios, video content is usually partitioned into small chunks of few seconds which can be transferred independently. Also, viewers only watch small parts of a video before leaving the session as stated in \cite{pattern} and \cite{pattern9}. This makes requesting the whole video from the CDN a waste in terms of cost and delivery latency. According to Ooyala’s Q4 2013 reports \cite{oolaya1}, the average watch time, per play, of VoD on mobiles is only 2.8 minutes. 
Ending a video unexpectedly without fully watching it incurs: (a) a lower cache hit ratio due to caching a number of unwatched video parts. This leads to a rapid cache storage saturation and a lower caching efficiency because of storing a long video in the same base station;  (b) a lower processing efficiency and a higher transcoding cost caused by generating videos that will not be fully watched, in addition to a rapid resource consumption because of transcoding a long video in one server. In the same context, authors in \cite{pattern9} showed that the watching time of videos can be studied since it is impacted by the length, the popularity and the category of a video.  Hence, based on these challenges, we propose to derive a study of chunks popularity based on users viewing pattern. Then, using this model, a proactive caching to load the cache with popular videos and a reactive cache replacement are introduced. We, also, propose a framework where servers can collaborate to cache or transcode chunks of the same video. This framework will be presented in a greedy heuristic.

The main contributions of this paper are summarized as follows: 
\begin{itemize}
	\item We present our collaborative chunks caching and processing. By proposing the possibility of caching one video in different cooperative MEC servers, resource utilization can be improved.   
	\item We propose a probabilistic model where we study the popularity of different chunks of videos based on users preference. 
	\item We introduce our popularity-aware caching policies (PcP and CrP) that use the probabilistic model for chunks caching and evicting.
	\item We design a PCCP greedy heuristic to schedule the chunks loading and transcoding. 
	\item We evaluated our model compared to previously described caching approaches (CachePro, CoCache and JCCP).
\end{itemize}
Our paper is organized as follows: In section \ref{system}, we present our proposed caching and processing collaborative system, where we study the viewing pattern of videos and we express the popularity of chunks. Then, the PcP and CrP policies are presented along with the PCCP greedy heuristic. A detailed experimental evaluation is provided in section  \ref{simulation}. Finally, in section \ref{conclusion}, we draw the conclusions.
\section{Proactive caching and processing of video chunks}\label{system}
In this section, we will describe our caching system implemented on distributed MEC servers in RAN networks. The description of the architecture is followed by the study of viewing pattern and the presentation of the probabilistic model. This model is used, next, to introduce the popularity-aware caching policies (PcP and CrP). Finally, the greedy heuristic suggesting a collaborative caching for different chunks is described.
\subsection{System model}
\begin{figure}[!h]
	\centering
	\includegraphics[scale=0.42]{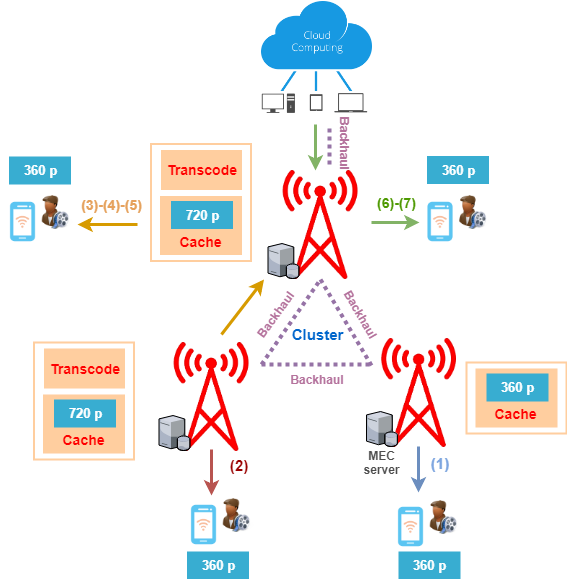}
	\caption{Illustration of the proposed collaborative chunks caching and processing system: (1) The chunk can be served from home server; (2) The chunk can be served from the home server after being transcoded; (2) The chunk can be served from the neighbor server; (3) The chunk can also be served and transcoded in the neighbor server; (4) The chunk can be served from the neighbor BS server and transcoded at the home server; (6)-(7) The chunk can be served from the cloud either directly or after being transcoded loacally. }
	\label{network}
\end{figure}

In our system, the network consists of multiple base stations (BSs), each BS is associated with a MEC server providing computation, storage and networking capacities. The area grouping communicating base stations is called \textit{cluster}. In this paper, we intend to use the servers for caching and computation and we assume that these servers can share resources. Then, the shared streams can be transmitted to mobile users if requested. In addition, the transcoder embedded with the server can transcode the shared video to the required bitrate version if needed. Hence, the requested data can be received either from the cache or the transcoder. A video transcoding is the lossy compression of a higher bitrate version to a lower version. The architecture of our system is described in Figure \ref{network}.
\begin{table*}[]
	\scriptsize 
	\centering
	\caption{Fit coefficients of PDF of the drop positions for different categories. }
	\label{fit}	
	\begin{tabular}{|l|l|l|l|l|l|l|l|l|l|l|l|}		
		\hline
		Category ($cg_y$) & $\alpha_y$ & $\beta_y$ & $\gamma_y$ & P-square & Category ($cg_y$) & $\alpha_y$ & $\beta_y$ & $\gamma_y$ & P-square  \\ \hline
		People & 2.39 & 0.56 & 0.0023 & 0.999 & Howto & 2.74 & 0.52&0.0153 & 0.999 \\ \hline
		Gaming & 1.98 & 0.45 & 0.0146& 0.999& Comedy & 2.89 &0.65&-0.0250 & 0.999 \\ \hline
		Entertainment & 2.41& 0.56& -0.0064 & 0.999 & Education & 2.40 & 0.54 & -0.0104& 0.999 \\ \hline
		News & 4.70 & 0.95 & -0.298&0.999 & Science & 2.53& 0.53 & 0.013 & 0.999  \\ \hline
		Music & 2.45& 0.51 &0.0178& 0.999 & Autos & 2.68& 0.58& 0.0016 & 0.999 \\ \hline
		Sports & 4.34& 0.92 &-0.267&0.999& Activism & 2.50& 0.59 & -0.0228 & 0.999\\ \hline
		Film & 2.32& 0.62& 0.0205&0.999& Pets & 3.089 & 0.69 & -0.066& 0.999 \\ \hline
	\end{tabular}
\end{table*}
In our system, a cluster can comprise $K$ base stations accommodating caching servers. A cluster is denoted by $\mathcal{K}=\{1,2,...K\}$. The collection of videos shared in the cluster is indexed by $\mathcal{V}=\{1,2,...V\}$. We assume that all requested videos have $\mathcal{M}$ bitrate versions. Each video can be partitioned into  chunks with similar length. The set of chunks related to a video $v$ is denoted as $v=\{v_1, v_2,...,v_i,..., v_c\}$, $c$ is the number of chunks in the video $v$. All video chunks having a bitrate version $l$ have the same size proportional to the bitrate, denoted as $s_l$. The set of all chunks that can be requested by a viewer is denoted as $\breve{\mathcal{V}}=\{v^l_i|\quad v_i\in v,\quad v\in \mathcal{V}, l=1,2,...\mathcal{M}\}$. We consider that a video chunk $v_i^l$ can be obtained by transcoding the video chunk $v_i^h$, if $l \leq h$, $\forall v_i\in v, v\in \mathcal{V}$ and $l, h \in \{1,2,...\mathcal{M}\}$. We consider that viewers can only request and receive videos from the closer BS, denoted as home node. In addition, we consider that each server  $k$ is provisioned with a cache capacity equal to  $S_k$ bytes. We model the video content caching by introducing the variable $C^{v_i^l}_{k} \in \{0,1\},$ $\forall \quad v_i\in v,$ $v\in \mathcal{V},$ $\forall \quad l  \in \{1...\mathcal{M}\},$ $\forall \quad k \in \mathcal{K}$, where  $C^{v_i^l}_{k}=1$, if $v_i^l$ is stored at the BS $k$ and $C^{v_i^l}_{k}=0$, if not. The cache capacity of a cache server associated to a BS $k$ is expressed by $\sum\limits_{v_i^l\in \breve{\mathcal{V}}} s_l . C^{v_i^l}_{k} \leq S_{k}$. Since transcoding videos is a computational intensive task and because of the real-time requirements of converting videos to the requested bitrate, we consider separate instances for each transcoding task, that are large enough to transcode the highest considered representation of videos. Let $P_k$ represent the processing capacity (number of transcoding instances) of the $k^{th}$ caching server. The description of different chunks fetching scenarios presented in Figure \ref{network} will be deferred to ulterior subsections. 
\subsection{Users viewing pattern}\label{section::proactive}
To study the viewing pattern, we will identify first the parameters that impact the preferences of users (video categories, length, popularity, etc). Then, we will define our viewing model and identify the most likely chunks to be requested within the collaborative cluster. In fact, authors in \cite{popularity} studied the characteristics of several types of videos and proved that the popularity $p_v$ of a video $v$ follows a Zipf distribution with a skew parameter $\alpha$. Studies  in \cite{popularity2} showed that videos popularities  depend strongly on its category. Additionally, authors stated that the popularity of different categories changes from a viewer to another. Another interesting conclusion is that the popularity of a category depends also on the location of the viewer. Other researchers studied the viewing pattern and the watching time of VoD contents, among them we can cite  \cite{pattern}. These studies concluded that viewers only watch small parts of a video before leaving the session and that short videos have higher probabilities to be fully watched. Other previous works (e.g. \cite{pattern9}) proved that in addition to its length, the popularity and the category of a video can impact its watching time. These conclusions motivate us to identify the probability that a user requests a specific chunk based on the popularity and the category of the video, and the watching time pattern.
\subsubsection{Probability of requesting a video}
We define a set $Cg=\{cg_1,...,cg_G\}$ of $G$ categories. We define, also, a set of users $U^j$ connected to a BS $j$ at the studied period. We define for each user $u^j_k$ a set $UP^j=\{p(cg_y|u^j_k)\quad|\quad \forall y=1..G,\forall k=1..K\}$, where $p(cg_y|u^j_k)$ is the probability  of requesting videos from a category $cg_y$ ($y \in \{1..G\}$) by the user $u^j_k$. The probability that a video belonging to a category $cg_y$ is requested by users $U^j$ connected to a BS $j$ can be calculated by summing the probabilities that $cg_y$ is selected by different users: 
{\small \begin{equation}
\begin{tabular}{l}
$p_j(cg_y)=\sum_{k=1}^{|U_j|}p(u^j_k)p(cg_y|u^j_k),$  
\end{tabular}
\end{equation} }
where $|U_j|$ is the number of users connected to the BS $j$ and $p(u^j_k)$ is the probability that a user $u^j_k$ requests a video. We assume that all viewers have the same probability to request a video, $p(u^j_1)=...=p(u^j_{|U_j|})=\dfrac{1}{|U_j|}$. Hence, the probability of requesting a video belonging to a category $cg_y$ within the BS $j$ can be expressed as follows:
{\small \begin{equation}
	\label{2}
\begin{tabular}{l}
$p_j(cg_y)=\dfrac{1}{|U_j|}\sum_{k=1}^{|U_j|}p(cg_y|u^j_k).$  
\end{tabular}
\end{equation} } 
Next, we will identify the probability of requesting a video $v$ belonging to a category $y$, namely $P_{j,y,v}$. First, let $p(cg_y,v)$ be the probability of choosing a video $v$ from videos inside a category $cg_y$. This probability can be written as follows:
{\small \begin{align}
	\label{1}
p_j(cg_y,v)=
\left\{
\begin{array}{ll}
\dfrac{p_v}{\sum_{i=1}^{V}p_i \times I_i^{cg_y} },   &\text{if $I_v^{cg_y}=1$.} \\
0, & \text{otherwise}.
\end{array}
\right.
\end{align}}
where $V$ is the total number of videos, $p_v$ is the popularity of $v$ inside the library  and $I_v^{cg_y}$ is a binary variable which indicates if $v$ belongs to $cg_y$ or not. The difference between the probabilities $p_{j_1}(cg_y,v_i)$ and $p_{j_2}(cg_y,v_i)$ is the popularity of videos within the BSs $j_1$ and  $j_2$. The probability of requesting a video $v$ belonging to a category $cg_y$ from the whole library can be expressed as follows:
{\small \begin{equation}
\begin{tabular}{l}
$P_{j,y,v}=p_j(cg_y) \times p_{j}(cg_y,v)$  
\end{tabular}
\end{equation} }
The next step is to calculate the probabilities $p_{j,y,v,v_i}$ of requesting chunks $v_i$ of a video $v$. As we stated previously, the watching time and the behavior of viewers depend on the length, category and popularity of the requested videos and this viewing pattern has different distributions depending on the studied data \cite{pattern9}. Hence, without loss of generality, we will first analyze and extract the viewers behavior based on a specific dataset \cite{dataset}. Any other data can be used to identify the viewing behavior. Also, we suppose that we study a population that has the same behavior of chunks viewing.
\subsubsection{Probability of requesting a chunk of video}
To study the viewers behavior and model the video watching pattern, we will use a real-life video dataset. Specifically, we will use video logs extracted from one of the most popular video providers "YouTube" collected by authors in \cite{dataset} and named \textit{TweetedVideos}. This dataset contains more than 5 Million videos published in July and August, 2016 from more than one million channels. The dataset contains also several metadata of videos including view count which gives an idea about the popularity of videos, the watch time and video duration.
\begin{figure*}
	\centering
	\begin{minipage}{.5\textwidth}
		\centering
		\includegraphics[scale=0.32]{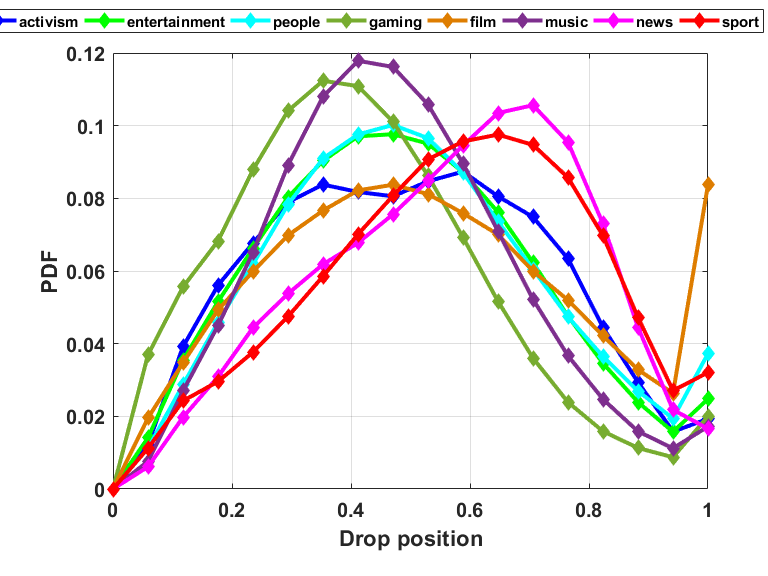}
		\caption{Dropping behavior modeling.}
		\label{all}
	\end{minipage}%
	\begin{minipage}{.5\textwidth}
		\centering
		\includegraphics[scale=0.25]{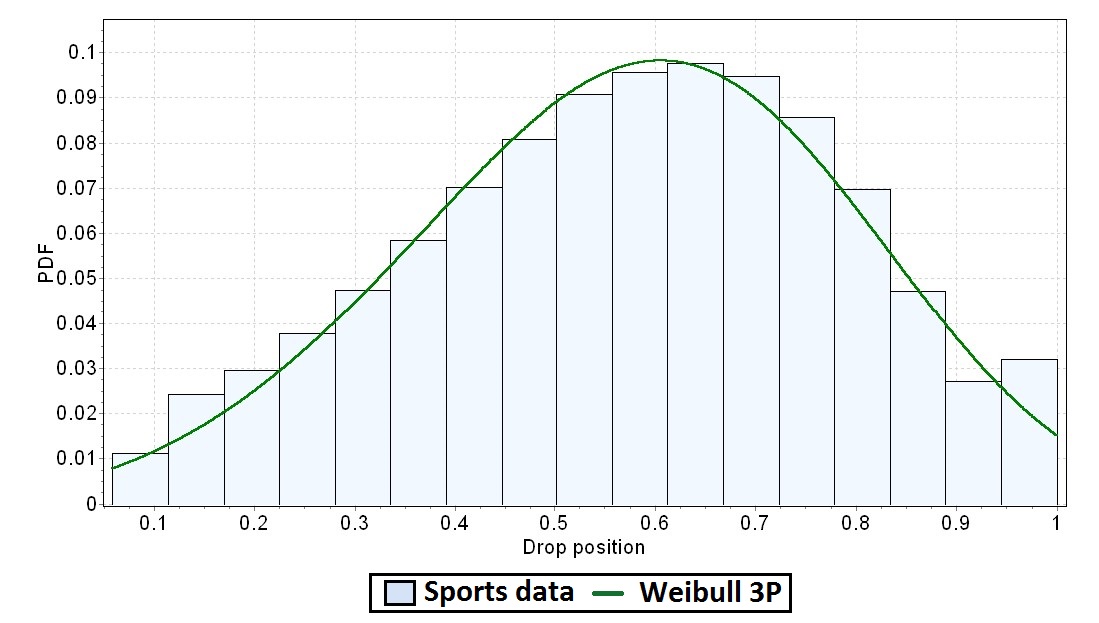}
		\caption{Drop position distribution fit.}
		\label{gaming_fit}
	\end{minipage}
\end{figure*}
In this paper, we suppose that the users request chunks of videos from the edge servers and different chunks have similar duration.  Hence, we decompose each video $v$ into chunks and we attribute to each chunk a position $v_i$. These positions are normalized from 0 to 1. 
We further model the drop position distribution which means the chunk position where the viewers stopped watching the video and the drop probability in each chunk. Figure \ref{all} presents the PDF distribution of the drop positions $Dp$ of videos from 14 different categories in the dataset. We can see that the drop position PDFs follow the same shape which is similar to a Weibull model but with different coefficients. This validates that the category of video impacts the viewing behavior. Figure \ref{gaming_fit} shows the integrity of our Weibull fit expressed as follows for a category $cg_y$:
 \begin{equation}
\begin{tabular}{l}
$Dp_y(c)=\dfrac{\alpha_y}{\beta_y}(\dfrac{c-\gamma_y}{\beta_y})^{\alpha_y-1}e^{-(\dfrac{c-\gamma_y}{\beta_y})^{\alpha_y}}$ 
\end{tabular}
\end{equation}
where $Dp_y(c)$ is the drop distribution of chunk position $c$ of a video belonging to a category $cg_y$.  $\alpha_y$, $\beta_y$ and $\gamma_y$ are the coefficients of the Weibull distribution; $\alpha_y>0$, $\beta_y>0$ and $-\infty<\gamma_y<+\infty$. Different Weibull coefficients corresponding to different categories of the dataset \textit{TweetedVideos} are presented in Table \ref{fit}. This table includes also the P-square indicators which is the goodness of fit measure. We can see that the P-square is close to 1 for all categories which proves that the Weibull fit can model the distribution of the drop positions.

Using the Weibull distribution, we can now derive the instantaneous drop probability and the watching probability of a random chunk position $v_i$. To simplify our model, we suppose that viewers leaving a video must have watched the last chunks completely. It means a viewer who dropped a video $v$ at a chunk $v_i$ must have watched this chunk. Hence, the probability of watching a chunk $v_i$ can be expressed as follows:
{\small \begin{equation}
\begin{tabular}{l}
$ p_v(v_i)= \dfrac{1-\int_{c=0}^{c=v_i}Dp_y dc}{\int_{c=0}^{c=1}Dp_y dc}= \dfrac{\int_{c=v_i}^{c=1}Dp_y dc}{\int_{c=0}^{c=1}Dp_y dc}.$\\ \\$= \int_{c=v_i}^{c=1}Dp_y dc.$\\ \\ 
$=\quad\Big[e^{-(\dfrac{v_i-\gamma_y}{\beta_y})^{\alpha_y}}-e^{-(\dfrac{1-\gamma_y}{\beta_y})^{\alpha_y}}\Big].$
\end{tabular}
\end{equation}}
where $1-\int_{c=0}^{c=v_i}D_p(v_i)$ is the probability of watching a video from position 0 to $v_i$ (not leaving a video between positions 0 and $v_i$) and $\int_{c=0}^{c=1}Dp_y dc$ is the probability of watching all chunks which is equal to 1. We define, now,  the popularity of a chunk $v_i$ of a video $v$ belonging to a category $cg_y$ as the probability of requesting a video $v$ and consuming its chunk $v_i$. From equations (\ref{2}) and (\ref{1}), we can estimate this popularity as:
{\small \begin{equation}
\begin{tabular}{l}
$P_{j,y,v,v_i}=P_{j,y,v} \times p_v(v_i).$\\ \\$=p_j(cg_y) \times p_{j}(cg_y,v)\times p_v(v_i).$
\end{tabular}
\end{equation}}
\newline
After deriving the expression of chunk popularity, we will define our popularity-aware policies in the next section. 
\subsection{Popularity-aware caching policies}\label{section::pref_aware}
\begin{figure}[!h]
	\centering
	\includegraphics[scale=0.45]{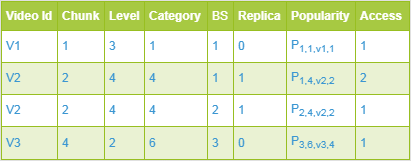}
	\caption{Snapshot of a synchronized catalogue.}
	\label{catalogue}
\end{figure}

In this section, we present two caching policies based on the popularity of chunks studied in the previous subsection; Proactive caching Policy PcP for cache pre-loading and Cache replacement Policy CrP for reactive cache removal.
However, we need to introduce, first, our proposed distributed and synchronized video catalogue that contains the metadata of each stored video and helps to manage the data caching. This catalogue is associated with each cache. In fact, each catalogue is updated reactively for every new event e.g., accessing, caching or removing occurring in the related base station. This updated catalogue is shared in real time with other nodes so they can update their catalogues accordingly. Such a collaborative catalogue management helps to enhance the video selection delay by making video searching local and minimizing the communication overheads. A snapshot of a synchronized catalogue is illustrated in Figure \ref{catalogue}. 
\subsubsection{CrP}
CrP is a reactive caching replacement policy that presents two new features: The first feature is the minimum replication strategy. Indeed, in previous edge caching studies, each fetched video is cached at the local cache whether it exists in the cluster or not, which means multiple copies of the same video can be cached. Even when the cache is full, other potentially needed videos have to be deleted to provide space for the duplicated video. In our system, a second copy of the video is only stored when the cache is available and this video is marked in the catalogue as \textit{Replica=1}. In case of storage unavailability, only a single copy of the video is cached in the collaborative cluster. In our system, a video chunk is called a \textit{Replica}, if a similar or a higher bitrate representation exists in the cluster because it is always possible to create a lower video quality from a higher one. In this way, when the cache is unavailable, if a higher representations exists, there is no need to store lower ones. The second new feature is the popularity-aware caching and removing. Indeed, if the cache is available, the incoming chunk request is stored even if it is a replica or an unpopular content. In case the storage is unavailable, all $P_{j,y,v,v_i}$ in the catalogue are ranked and the Less Popular Chunk (LPC) is selected. If the requested video chunk is a replica and its popularity is less than LPC, this chunk is not cached. The above approach guarantees that the highest popular chunks are cached and the number of stored videos is maximized due to the minimum replication policy and the LPC rule. The detailed CrP policy is presented in algorithm \ref{RcP}.   
\begin{algorithm}[!h]
	\scriptsize
	\caption{CrP}
	\label{RcP}
	\begin{algorithmic}[1]
		\State \textbf{Input: $S_j, v_i^l, ctg, C_j$}, \textbf{Output:} updated $S_j, ctg, C_j$
		\State $isReplica=0$, $Add=0$
		\If {$v_i^l \in ctg$}  $isReplica=1$
		\EndIf
		\If {$S_j<s_l$}
		\State $LPC =min(P_{j,.,.,.})^*$		
		\If {$isReplica=1$}				
		\If {$LPC<P_{j,y,v,v_i}$}
		\State \underline{* RcP-Removing part:}
		\While {$S_j<s_l$}
		\State - Prioritize removing unpopular replicas 
		\State $v_{i1}^{l1}$ over  unique unpopular chunks
		\State $S_j=S_j+s_{l1}$, $C_j^{v_{i1}^{l1}}=0$
		\EndWhile
		\State $Add=1$
		\EndIf
		\Else				
		\If {$LPC>P_{j,y,v,v_i}$}
		\While {$S_j<s_l$}
		\State - Remove Higher popular chunks $v_{i2}^{l2}$ with \State$P_{j,y,v,v_i2}-LPC<TH$
		\State $S_j=S_j+s_{l2}$, $C_j^{v_{i2}^{l2}}=0$
		\EndWhile
		\State $Add=1$
		\Else
		\State \underline{* RcP-Removing part}
		\EndIf			
		\EndIf	
		\Else
		\State $Add=1$
		\EndIf		
		\If {$Add=1$}
		\State $S_j=S_j-s_l$, $C_j^{v_i^l}=1$
		\State - Add $v_i^l$ to $ctg$
		\State - Update $v_i^l$ to recent time in $ctg$
		\State - Update other catalogues
		\Else
		\State - Relay to the viewer without caching
		\EndIf
		\State \textbf{*} $P_{j,.,.,.}$ is the set of popularities of different chunks stored in the BS $j$
	\end{algorithmic}
\end{algorithm}
\subsubsection{PcP}
We, also, propose a proactive caching policy where we populate the cache with the most popular chunks that are most likely to be requested. This pre-load is done based on the preference of active users connected to the base station as studied previously in section \ref{section::proactive}. More specifically, high popular chunks are stored with the higher bitrate one by one until filling the cache. If the cache is still available, popular chunks are pre-loaded again with a lower representation. Meanwhile, the catalogue is updated with \textit{Replica} and popularity status. This task is done at the initialization of the network and has an initial cost. The proposed PcP is illustrated in algorithm \ref{PcP}. 

\begin{algorithm}[!h]
	\scriptsize
	\caption{PcP}
	\label{PcP}
	\begin{algorithmic}[1]
		\State \textbf{Input: $S_1,...,S_K, P_{1,.,.,.},...,P_{K,.,.,.}$}
		\State \textbf{Output: $C_1, ..., C_K$, $ctg$, $S_1,...,S_K$ }
		\State \textbf{Cache initialization: }$C_1=0, ..., C_K=0$
		\State \textbf{Catalogue initialization: }$ctg=\varnothing$
		\State level=$\mathcal{M}$ 
		\For {$j \in {\{1,..,K\}}$}		
		\While {$S_j>0$}
		\State - Select the higher popular chunk $max(P_{j,.,.,.})$
		\State - Update $ctg$
		\State $S_j=S_j-s_{level}$, $C_j^{v_{i}^{level}}=1$
		\If {All chunks are cached}
		\State level=level-1
		\EndIf
		\EndWhile
		\EndFor
	\end{algorithmic}
\end{algorithm}
\subsection{Proactive Chunks Caching and Processing (PCCP)}\label{sec::algorithm}
To illustrate the different events that can occur when a user connected to a BS $j$ requests a chunk of video $v_i^l$, we introduce our greedy algorithm, named Proactive Chunks Caching and Processing (PCCP) and detailed in algorithm \ref{heuristic}.
At the installation of the system, different caches and catalogues are initialized with the popular video chunks. On each chunk request arriving to the BS $j$, the catalogue is checked to find the chunk with the requested bitrate. If available, the viewer is served from the home node $j$ (see Figure \ref{network}, scenario 1) and the CrP policy is called to update the access time of the video. In addition to the requested chunk, $Wd$ potentially requested chunks from the same video are fetched. In this way, when the user requests to watch the next part,
the video will be streamed without stalls. If the requested bitrate is not available at home node, a higher representation is searched locally (see Figure \ref{network}, scenario 2) and the possibility of serving the video from a neighboring BS is also studied. The option that has the lower cost is adopted. If the requested bitrate does not exist in the cluster and a local transcoding is not possible, a higher representation is searched in neighbor nodes. Depending on resources availability, the chunk of video can be transcoded at the neighboring node or locally (see Figure \ref{network}, scenarios 3,4 and 5) . If these options cannot be performed, the request is served from the CDN, either by bringing the same or a higher bitrate (see Figure \ref{network}, scenario 6 and 7). The CrP policy is applied to update the cache $j$ and the catalogue. It means, depending on the popularity of the requested chunk, the caching and removing are accomplished.
\begin{figure*}[!h]
	\centering
	\mbox{
		\hspace{-0.5cm}
		\subfigure[\label{cachea}]{\includegraphics[scale=0.3]{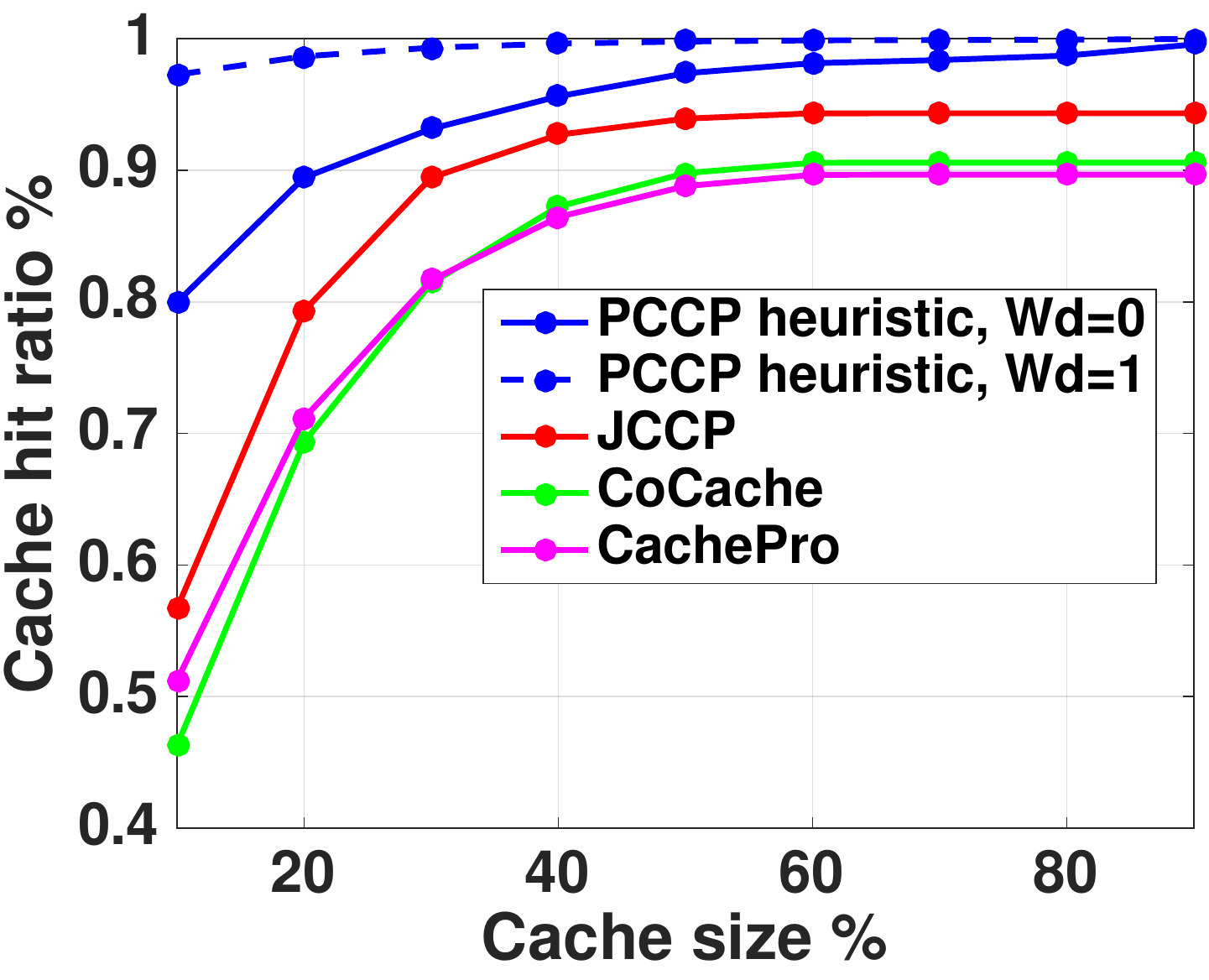}}
		\hspace{-0.1cm}
		\subfigure[\label{cacheb}]{\includegraphics[scale=0.3]{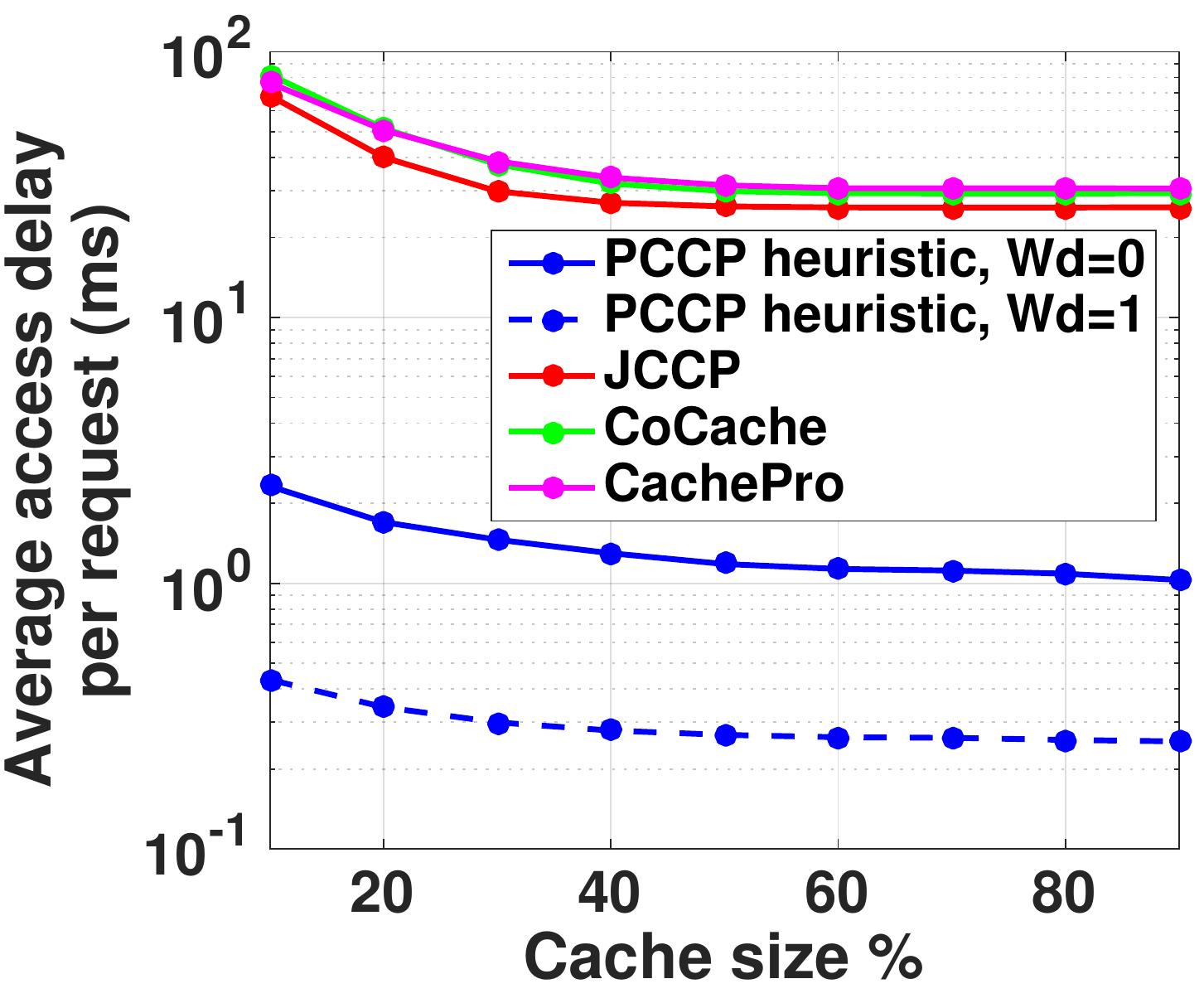}}
		\subfigure[\label{cachec}]{\includegraphics[scale=0.3]{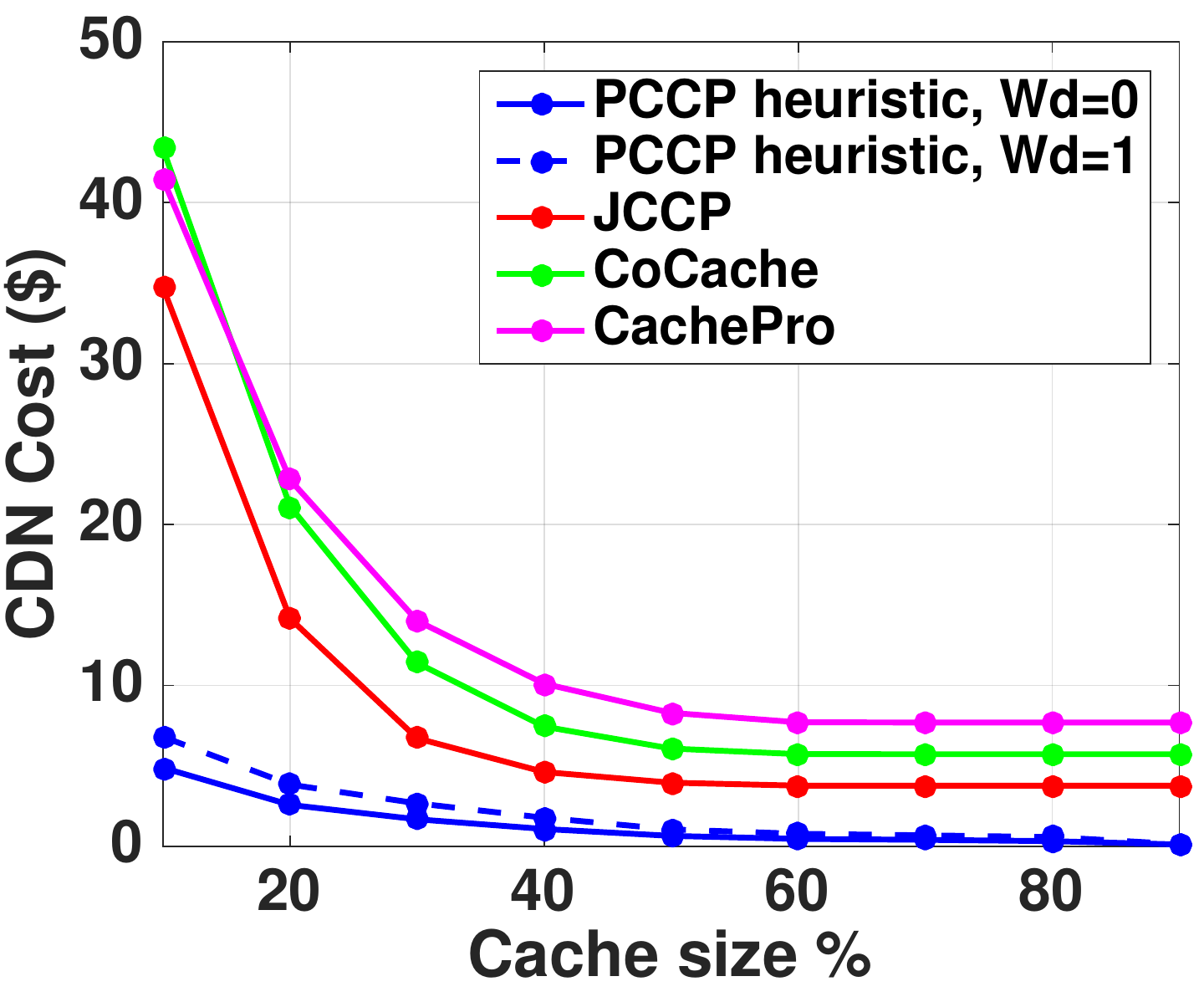}}	
	}\\
	\mbox{
		\hspace{-0.5cm}
		\subfigure[\label{proca}]{\includegraphics[scale=0.3]{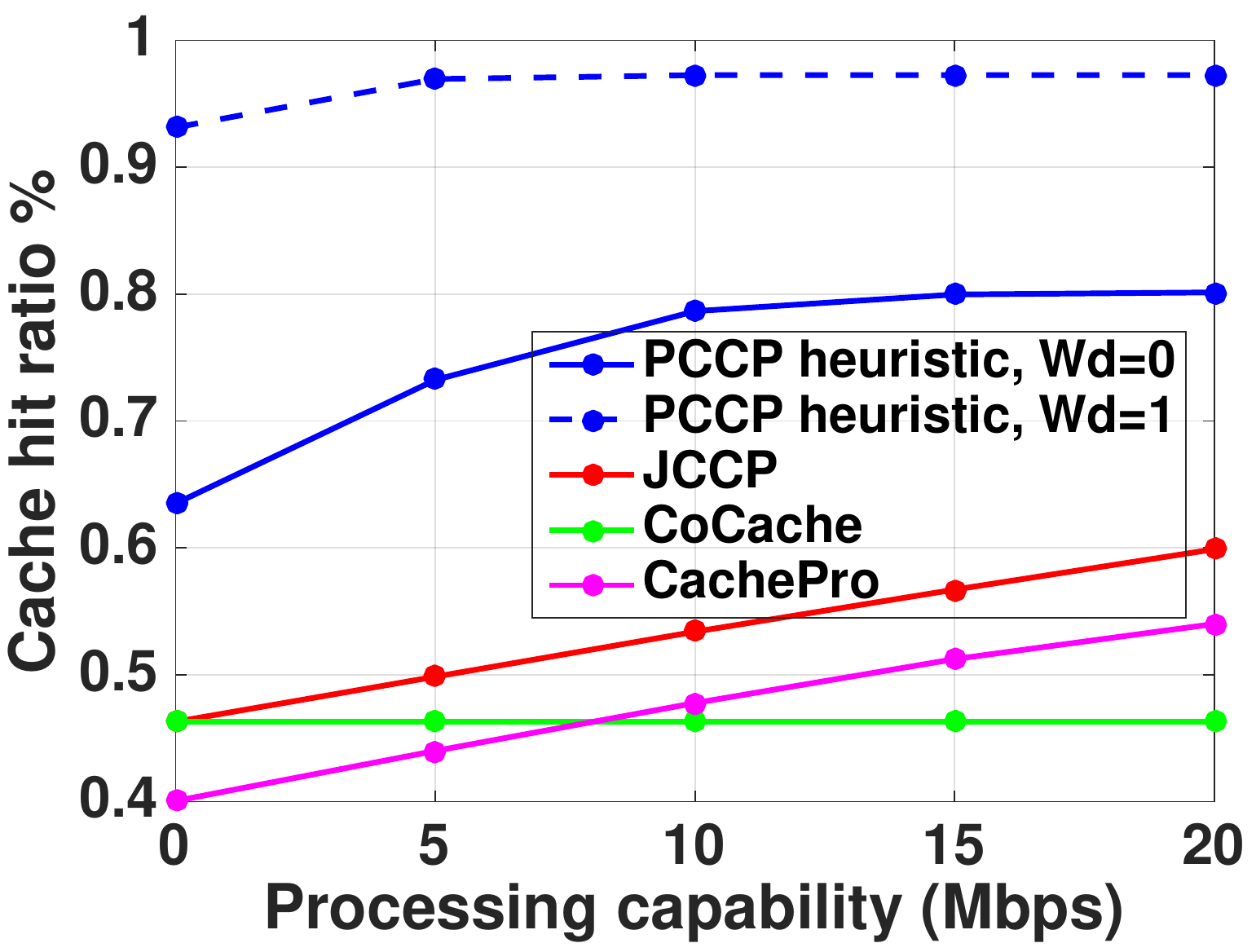}}
		\hspace{-0.3cm}
		\subfigure[\label{procb}]{\includegraphics[scale=0.3]{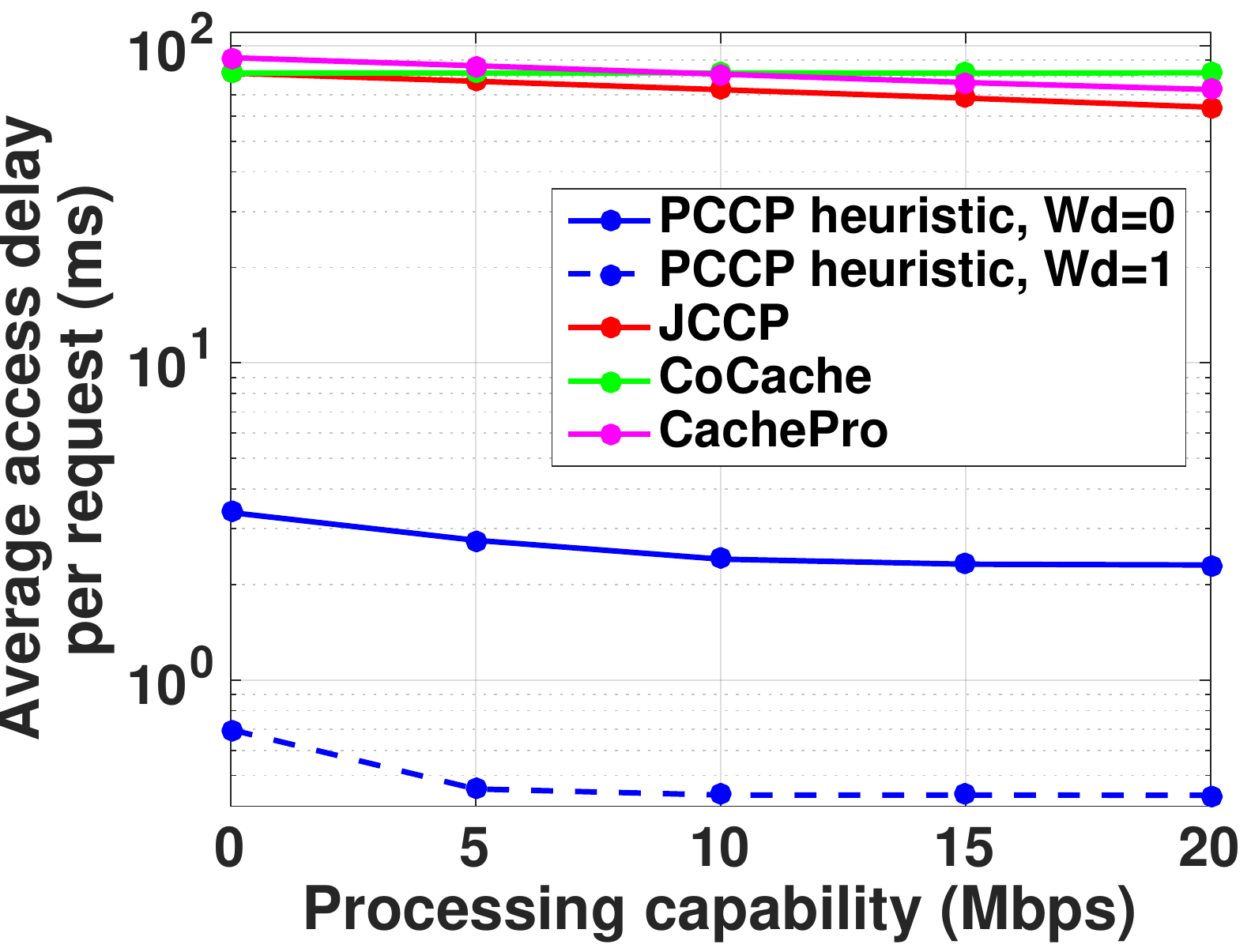}}
		\hspace{-0.3cm}	
		\subfigure[\label{procf}]{\includegraphics[scale=0.29]{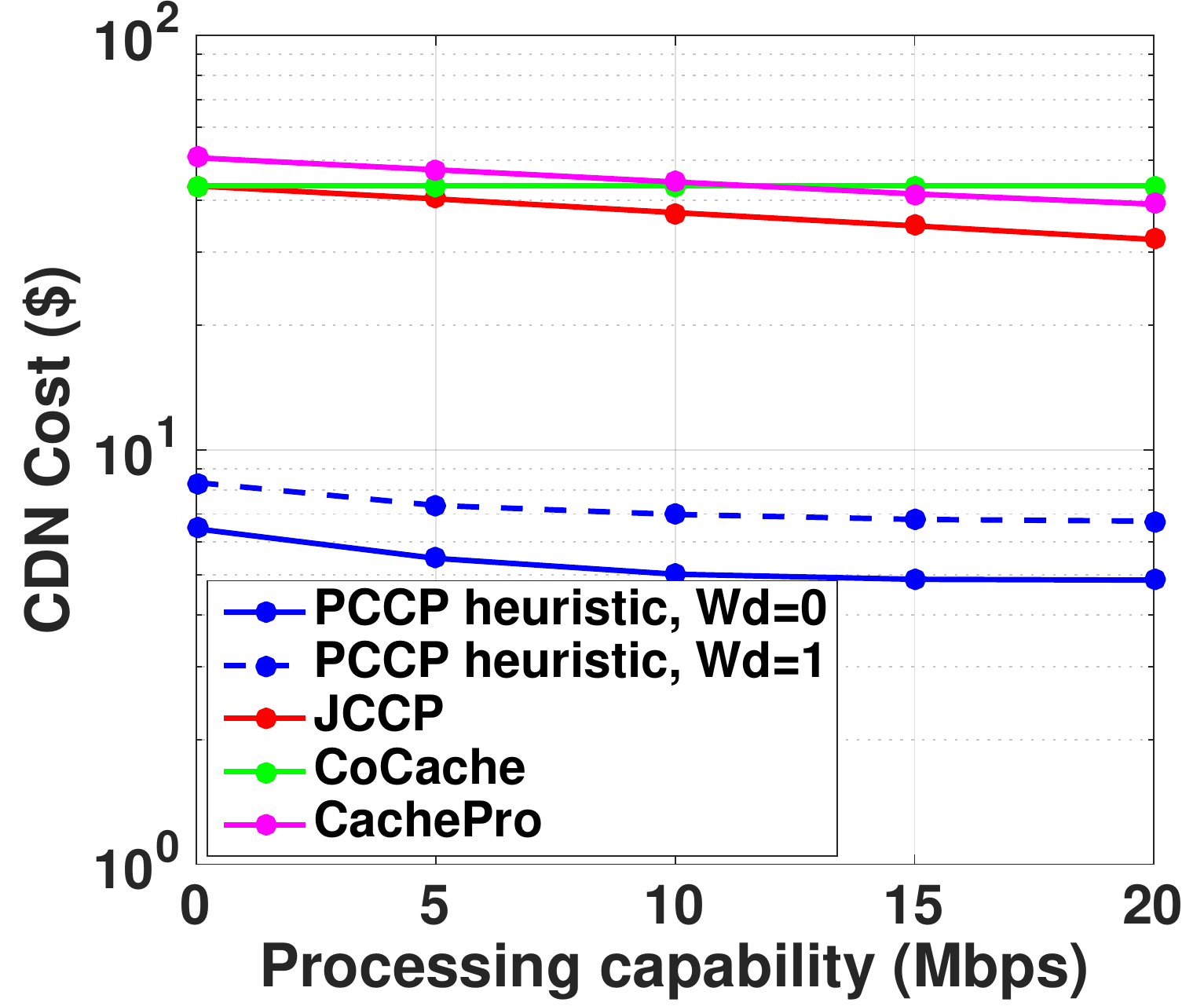}}
	}
	\caption{Performance comparison based on varying cache capacity and a processing capacity =15Mbps: (a) Cache Hit ratio (b) Average access delay per request, (c) CDN cost; Performance comparison based on varying processing capacity and a cache size =10\% of library:  (d) Cache Hit ratio, (e) Average access delay, (f) CDN cost.}
	\label{cache}
\end{figure*}
\begin{algorithm}[H]
	\scriptsize
	\caption{PCCP}
	\label{heuristic}
	\begin{algorithmic}[1]
		\State \textbf{Initialize}$P_{1,.,.,.},...,P_{K,.,.,.}$, $S_1, ..., S_K$
		\State ($C_1, ..., C_K$, $ctg$, $S_1,...,S_K$)=\textbf{PcP}($S_1, ..., S_K$, $P_{1,.,.,.},...,P_{K,.,.,.}$)
		\For {$j \in {\{1,..,K\}}$}
		\For {each video request $v_i^l$ incoming to BS $j$}
		\For {$w \in {\{0,..,Wd\}}$}	
		\If {$C^{v_{i+w}^l}_{j}=1$}	
			\State -Stream from home BS $j$
		\State -\textbf{CrP}($S_j, v_{i+w}^l, ctg, C_j$)
		\Else
		\If {$\sum\limits_{h=l+1}^{\mathcal{M}}C^{v_{i+w}^h}_{j}\geq1$ and $P_j >0$}
		\If {$C^{v_{i+w}^l}_{k}=1$, $k \in \mathcal{K}$}
		\If {$Cb^l>Ct^l$}
		\State -Transcode and Stream from 
		\State home BS $j$
		\State -\textbf{CrP}($S_j, v_{i+w}^l, ctg, C_j$)
		\Else
		\State -Fetch from neighboring BS $k$
		\State -\textbf{CrP}($S_j, v_{i+w}^l, ctg, C_j$)
		\EndIf
		\Else
		\State -Transcode and Stream from home BS $j$
		\State -\textbf{CrP}($S_j, v_{i+w}^l, ctg, C_j$)
		\EndIf 	  
		\Else
		\If {$C^{v_{i+w}^l}_{k}=1$, $k \in \mathcal{K}$}
		\State -Fetch from neighboring BS $k$
		\State -\textbf{CrP}($S_j, v_{i+w}^l, ctg, C_j$)
		\Else
		\If {$\sum\limits_{h=l+1}^{\mathcal{M}}C^{v_{i+w}^h}_{k}\geq1,
			k \in \mathcal{K}$ \\ $\qquad \qquad \qquad \qquad \quad$ and $P_k >0$}
		\If {$P_j>P_k$} 
		\State -Fetch from $k$ and transcode at BS $j$
		\State -\textbf{CrP}($S_j, v_{i+w}^l, ctg, C_j$) 
		\State -\textbf{CrP}($S_j, v_{i+w}^h, ctg, C_j$)	
		\Else
		\State -Transcode and Fetch from node $k$
		\State -\textbf{CrP}($S_j, v_{i+w}^l, ctg, C_j$)		
		\EndIf
		\Else 
			\If {$P_j>0$}
		\State -Fetch $v_{i+w}^{\mathcal{M}}$ from CDN and 
		\State transcode at home BS $j$			
		\State -\textbf{CrP}($S_j, v_{i+w}^{\mathcal{M}}, ctg, C_j$)
		\State -\textbf{CrP}($S_j, v_{i+w}^l, ctg, C_j$)
		\Else
		\State -Fetch $v_{i+w}^l$ from CDN 
		\State -\textbf{CrP}($S_j, v_{i+w}^l, ctg, C_j$) 
		\EndIf
		\EndIf
		\EndIf
		\EndIf
		\EndIf 
		\EndFor
		\EndFor
		\EndFor	
	\end{algorithmic}
\end{algorithm}
%
\section{Performance evaluation}\label{simulation}

\subsection{Simulation Settings}
In this section, we will evaluate the performance of our system under different network configurations including storage capacity, processing capacity and popularity of videos. In our simulation, we considered that our network consists of 3 neighboring BSs attached to 3 MEC servers ($\mathcal{K}=3$). Let the video library $\mathcal{V}$ consist of 1000 different videos chosen randomly from the dataset in \cite{dataset} and belonging to $G=$ 14 categories described previously in Table \ref{fit}.  All videos are divided into chunks of 30 seconds each. We selected only videos with a duration lower than 1500 s (50 chunks) to limit the size of the set $P_{j,.,.,.}$. Each video has 4 different representations ($\mathcal{M}=4$). As configured in \cite{joint2}, we will set the bitrate variants related to each representation to be 0.45, 0.55, 0.67, 0.82 of the original video bitrate version. In this paper, we consider that all videos have the same original bitrate which is 2 Mbps. The popularity (number of views) of the chosen videos follows a Zipf distribution having a skew parameter $\alpha=0.5$. The parameters of users arrival and requests along with the cluster parameters are summarized in Table \ref{eval_parameters}.
\begin{table}[!h]
	\caption{Simulation parameters.}
	\label{eval_parameters}
	\begin{tabular}{|l|l|}
		\hline
		Variable & Distribution/parameters value \\ \hline
		Number of MEC servers & const, $\mathcal{K}$=3\\ \hline
		Total number of video requests & const, $R_1=R_2=R_3=$10.000 \\ \hline
		Total number of videos & const, $\mathcal{V}$=1000 randomly chosen\\ &from \cite{dataset} \\ \hline
		Video popularity & zipf, $\alpha$=0.5 \\ \hline
		Number of video categories & const, $G=$14 (see Table \ref{fit}) \\ \hline
		Category preference $UP^j$ & random \\ \hline
		Video sizes & $\leq$1500s (50 chunks) \\ \hline
		Video bitrate & Uniform, $\mathcal{M}$=4, from 200 kbps \\ &to 2Mbps \\ \hline
		Watching time & Exponential, mean watch time\\ &from \cite{dataset} \\ \hline
		Chunk size & 30s \\ \hline
		Number of viewers & const, $|U_1|=|U_2|=|U_3|=$ 500 \\ \hline
		Activity session size & Exponential, mean 300s \\ \hline
		Video request arrival & Poisson, $\lambda$=5, inter-arrival time=30s \\ \hline
		Max cache size & Library size \\ \hline
		popularity threshold  & $TH$=0.001 \\ \hline		
	\end{tabular}
\end{table}
\newline
The latency of getting a chunk of video for various scenarios follows a uniform distribution in a range of (a) [5-10] ms from the cloud remote servers  (b) [1-2.5] ms when fetching different versions from the neighboring BSs (c) [0.25-0.5] ms from the home server. Several parameters are evaluated to prove the performance of our system: (a) \textbf{cache hit ratio:} is the number of requests that can be fetched or transcoded in the edge network (home or neighbor servers). (b) \textbf{Average access delay:} is the average latency to receive videos from different caches or from the CDN. (d) \textbf{CDN cost:} is the cost of fetching the chunks of videos from the CDN. The CDN cost is calculated as 0.03\$ per GB. Our system is compared to different recently proposed systems, which are CachePro\cite{cache1}, CoCache\cite{cache5}, and JCCP\cite{joint2}.
\subsection{Simulation Results}
\subsubsection{Impact of caching and processing resources}
The cache size and the processing capacity are important parameters to test the efficiency of a caching system. Figure \ref{cache} shows the performance of different cache systems achieved for the described cluster for different cache and processing sizes. The results show the performance of our system in terms of cache hit ratio, access delay and cost. It can be seen that our PCCP heuristic with its CrP and PcP policies performs significantly better than the other caching systems even for low cache sizes or processing capability. For example, when the cache size is equal to 10\% of the library size, PCCP achieves a hit ratio equal to 0.79 when $Wd=0$ and 0.97 when $Wd=1$ compared to JCCP, CoCache and CachePro achieving a cache hit ratio equal to 0.56, 0,46 and 0.51 respectively. We can see that our system performs more than 20 \% better than the other systems without proactively fetching the next chunks ($Wd=0$). This can be explained by: (a) The proactive caching policy (PcP) that stores the highest popular chunks, which improves the probability to find the videos inside the cluster, (b) Collaboration between BSs to store different chunks of the same videos which makes caching a video possible, even when the cache is full, (c) Avoiding to store the whole video, since it is proved that viewers rarely watch the content to the end, which provides more space to cache a higher number of chunks, (d) The reactive caching policy (CrP) that stores only high popular chunks in the cache, removes only less popular chunks and avoids the replication of videos to maximize the number of cached videos. When $Wd=1$, we can see that the hit ratio becomes very high, which is explained by the fact that the next potentially requested chunks are fetched and served beforehand. Such prediction of the incoming requests contributes to increasing the hit ratio. The system with a prediction window $Wd=1$ incurs more cost and data fetching compared to the system without prediction. This can be explained by the additional cost of  the chunk that is served but not watched by the user abandoning the video. This cost waste can be accepted since the loaded content is only a small chunk with a minimal cost.
\subsubsection{Impact of video popularity}
Varying the popularities of videos has an important impact on our system since it is based on the preference of viewers.
\begin{figure}[H]
	\centering
	\mbox{
	\hspace{-0.5cm}
	\subfigure[\label{zipfa}]{\includegraphics[scale=0.29]{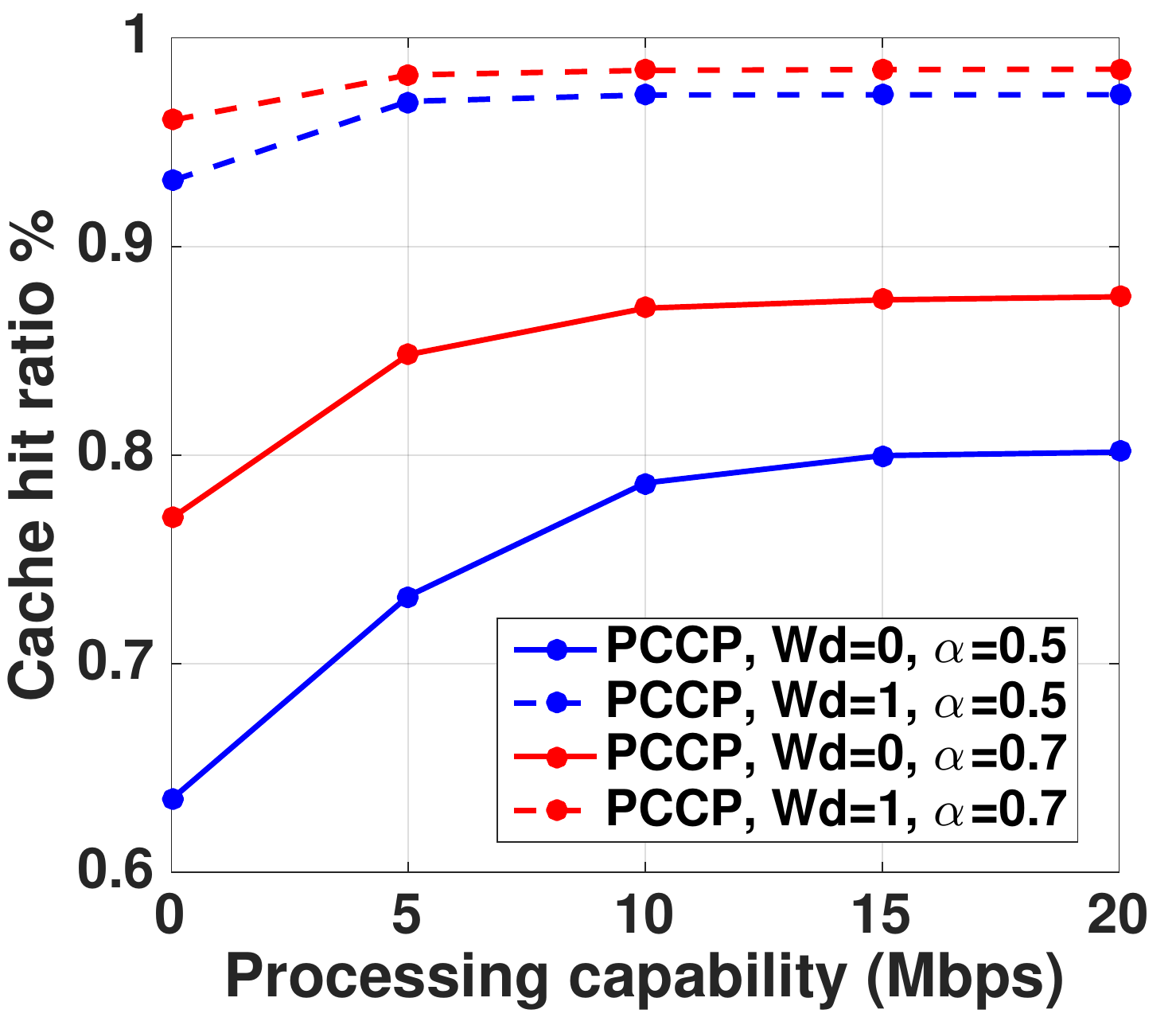}}	
	\subfigure[\label{zipfc}]{\includegraphics[scale=0.29]{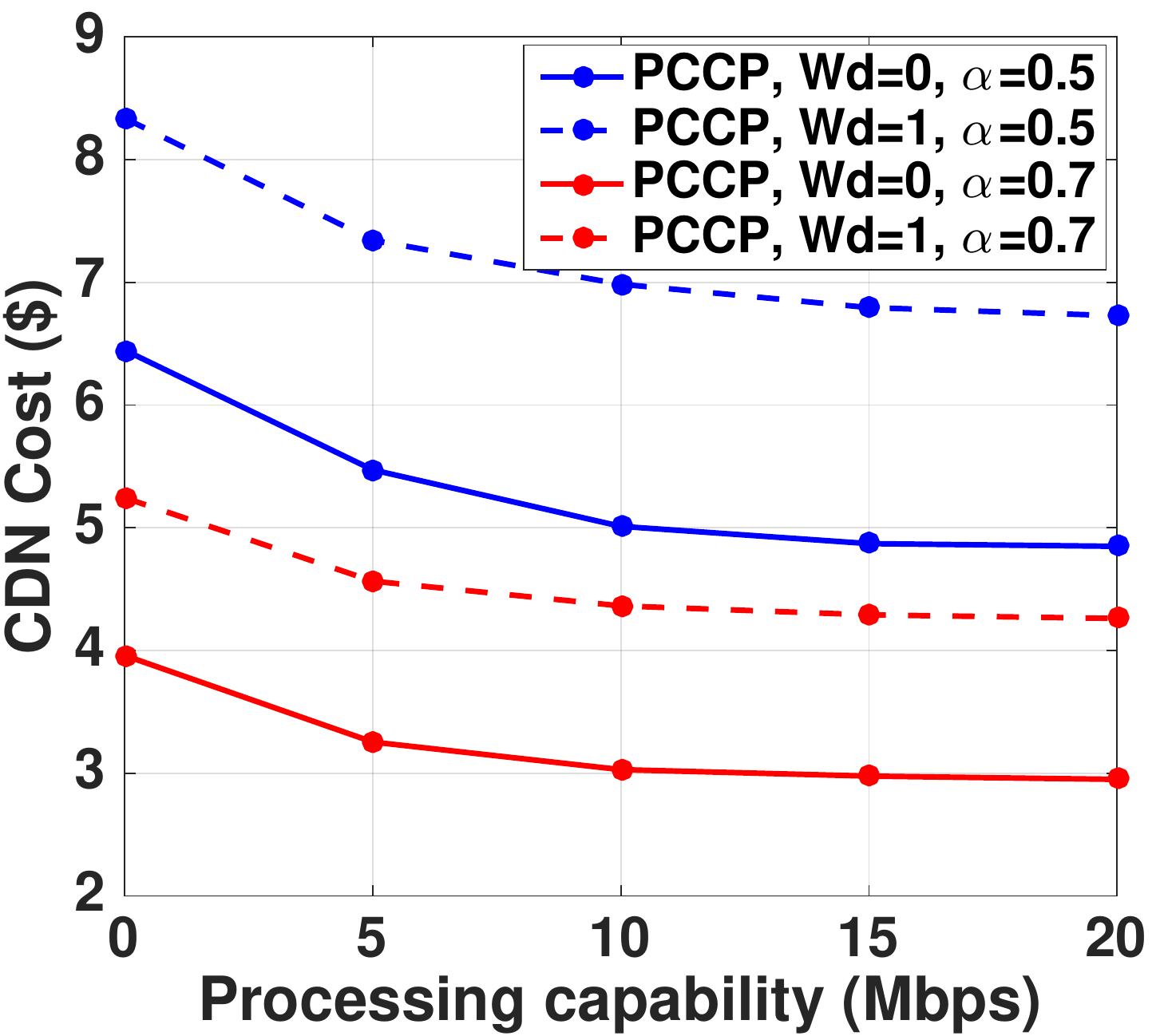}}	
   }
	\caption{Performance comparison based on Zipf parameter with a cache size = 10\% of the library size : (a) Cache Hit ratio, (c) CDN cost.}
	\label{zipf}
\end{figure}
 Figure \ref{zipf} presents the impact of changing the Zipf parameter $\alpha$ on cache hit and CDN cost. Specifically, a large $\alpha$ (high skewed popularity) represents a library with a small number of videos with similar popularities. It means the library contains videos with very high popularities and videos with very low popularities. In our simulation, we created another video library from the same dataset \cite{dataset} containing videos with a larger $\alpha=0.7$, which makes the library contain videos with higher difference in terms of popularity. In this way, videos with very high popularity will be highly requested. Whereas, unpopular videos will be rarely or never requested. Figures \ref{zipfa} and \ref{zipfc} show that a higher skewed library gives better results in terms of cache ratio. This is explained by the fact that only a part of videos are frequently requested which are stored in the cluster thanks to the PcP policy. Also, even if the cache is low (10\% of the library size), the size is enough to cache highly popular chunks which enhances the cache hit. 
\section{Conclusion}\label{conclusion}
In this paper, we propose that different MEC servers collaborate to cache and transcode different chunks of one video content. In order to maximize the edge caching, we studied videos viewing pattern and we proposed CrP and PcP content placement policies for estimating the placement of video chunks in the BS caches based on a probabilistic model. Then, to schedule sharing videos between MEC servers, a greedy algorithm (PCCP) is proposed. The extensive simulation of our heuristic proves the performance of PCCP compared to other caching approaches in terms of cost, cache hit ratio and access delay.
\section*{Acknowledgment}
This publication was made possible by NPRP grant 8-519-1-108 from the Qatar National Research Fund (a member of Qatar Foundation). The findings achieved herein are solely the responsibility of the author(s).
\bibliographystyle{IEEEtran}
\bibliography{chunks_conference}

\begin{thebibliography}{10}
\providecommand{\url}[1]{#1}
\csname url@samestyle\endcsname
\providecommand{\newblock}{\relax}
\providecommand{\bibinfo}[2]{#2}
\providecommand{\BIBentrySTDinterwordspacing}{\spaceskip=0pt\relax}
\providecommand{\BIBentryALTinterwordstretchfactor}{4}
\providecommand{\BIBentryALTinterwordspacing}{\spaceskip=\fontdimen2\font plus
\BIBentryALTinterwordstretchfactor\fontdimen3\font minus
  \fontdimen4\font\relax}
\providecommand{\BIBforeignlanguage}[2]{{%
\expandafter\ifx\csname l@#1\endcsname\relax
\typeout{** WARNING: IEEEtran.bst: No hyphenation pattern has been}%
\typeout{** loaded for the language `#1'. Using the pattern for}%
\typeout{** the default language instead.}%
\else
\language=\csname l@#1\endcsname
\fi
#2}}
\providecommand{\BIBdecl}{\relax}
\BIBdecl

\bibitem{statistics2016}
\BIBentryALTinterwordspacing
(2016) Global internet phenomena report. [Online]. Available:
  \url{https://www.sandvine.com/trends/global-internet-phenomena/}
\BIBentrySTDinterwordspacing

\bibitem{mobileData}
``Cisco visual networking index: Global mobile data traffic forecast update,
  2016–-2021,'' in \emph{Cisco white paper}, 2017.

\bibitem{cache1}
H.~Ahlehagh and S.~Dey, ``Video-aware scheduling and caching in the radio
  access network,'' \emph{IEEE/ACM Transactions on Networking}, vol.~22, no.~5,
  pp. 1444--1462, Oct 2014.

\bibitem{cache5}
J.~Dai, F.~Liu, B.~Li, B.~Li, and J.~Liu, ``Collaborative caching in wireless
  video streaming through resource auctions,'' \emph{IEEE Journal on Selected
  Areas in Communications}, vol.~30, no.~2, pp. 458--466, 2012.

\bibitem{joint2}
T.~X. Tran, P.~Pandey, A.~Hajisami, and D.~Pompili, ``Collaborative
  multi-bitrate video caching and processing in mobile-edge computing
  networks,'' \emph{CoRR}, 2016.

\bibitem{pattern}
H.~Yu, D.~Zheng, B.~Y. Zhao, and W.~Zheng, ``Understanding user behavior in
  large-scale video-on-demand systems,'' \emph{SIGOPS Oper. Syst. Rev.},
  vol.~40, no.~4, pp. 333--344, 2006.

\bibitem{pattern9}
Y.~Chen, B.~Zhang, Y.~Liu, and W.~Zhu, ``Measurement and modeling of video
  watching time in a large-scale internet video-on-demand system,'' \emph{IEEE
  Transactions on Multimedia}, vol.~15, no.~8, pp. 2087--2098, 2013.

\bibitem{oolaya1}
\BIBentryALTinterwordspacing
(2013) Ooyala’s q4 2013 report. [Online]. Available:
  \url{http://www.ooyala.com/resources/industry-reports}
\BIBentrySTDinterwordspacing

\bibitem{popularity}
M.~Cha, H.~Kwak, P.~Rodriguez, Y.~Y. Ahn, and S.~Moon, ``Analyzing the video
  popularity characteristics of large-scale user generated content systems,''
  \emph{IEEE/ACM Transactions on Networking}, vol.~17, no.~5, pp. 1357--1370,
  2009.

\bibitem{popularity2}
Z.~Li, J.~Lin, M.-I. Akodjenou, G.~Xie, M.~A. Kaafar, Y.~Jin, and G.~Peng,
  ``Watching videos from everywhere: A study of the pptv mobile vod system,''
  in \emph{Proceedings of the 2012 Internet Measurement Conference}, ser. IMC
  '12.\hskip 1em plus 0.5em minus 0.4em\relax ACM, 2012, pp. 185--198.

\bibitem{dataset}
S.~Wu, M.~Rizoiu, and L.~Xie, ``Beyond views: Measuring and predicting
  engagement on youtube videos,'' \emph{CoRR}, vol. abs/1709.02541, 2017.

\end{thebibliography}
\end{document}